\documentclass{PoS}
\let\ifpdf\relax
\usepackage[authoryear,square]{natbib}
\bibpunct{(}{)}{;}{a}{}{,}

\title{Pulsars in Globular Clusters with the SKA}

\ShortTitle{Pulsars in GCs}

\author{\speaker{J.~W.~T. Hessels}$^{ab}$, A. Possenti$^{c}$, M. Bailes$^{de}$, C.~G. Bassa$^{a}$, P.~C.~C. Freire$^{f}$, ~~~~~~~~~~~~~~D.~R. Lorimer$^{g}$, R. Lynch$^{h}$, S.~M. Ransom$^{i}$ \& I.~H. Stairs$^{j}$\\
       \llap{$^a$}ASTRON, the Netherlands Institute for Radio Astronomy, Postbus 2, 7990 AA, Dwingeloo, The Netherlands; E-mail: \email{hessels@astron.nl}\\
       \llap{$^b$} Anton Pannekoek Institute for Astronomy, University of Amsterdam, Science Park 904, 1098 XH Amsterdam, The Netherlands\\
       \llap{$^c$} INAF-Osservatorio Astronomico di Cagliari, Via della Scienza 5, I-09047 Selargius, Italy \\
       \llap{$^d$} Centre for Astrophysics and Supercomputing, Swinburne University of Technology, PO Box 218, VIC 3122, Australia\\
       \llap{$^e$} ARC Centre of Excellence for All-Sky Astrophysics (CAASTRO) \\
       \llap{$^f$} Max-Planck-Institut f{\"u}r Radioastronomie, auf dem H{\"u}gel 69, 53121, Bonn, Germany\\
       \llap{$^g$} Dept. of Physics and Astronomy, West Virginia University, Morgantown, WV 26506, USA\\
       \llap{$^h$} Dept. of Physics, McGill University, 3600 University Street, Montreal, QC H3A 2T8, Canada\\
       \llap{$^i$} National Radio Astronomy Observatory, 520 Edgemont Road, Charlottesville, VA 22901, USA\\
       \llap{$^j$} Dept. of Physics and Astronomy, University of British Columbia, 6224 Agricultural Road, Vancouver, BC V6T 1Z1, Canada\\
       }

\abstract{
Globular clusters are highly efficient radio pulsar factories. These pulsars can be used as precision probes of the clusters' structure, gas content, magnetic field, and formation history; some of them are also highly interesting in their own right because they probe exotic stellar evolution scenarios as well as the physics of dense matter, accretion, and gravity. Deep searches with SKA1-MID and SKA1-LOW will plausibly double to triple the known population. Such searches will only require one to a few tied-array beams, and can be done during early commissioning of the telescope --- before an all-sky pulsar survey using hundreds to thousands of tied-array beams is feasible. With SKA2 it will be possible to observe most of the active radio pulsars within a large fraction of the Galactic globular clusters, an estimated population of $600 - 3700$ observable pulsars (those beamed towards us). This rivals the total population of millisecond pulsars that can be found in the Galactic field; fully characterizing it will provide the best-possible physical laboratories as well as a rich dynamical history of the Galactic globular cluster system.
}

\FullConference{
Advancing Astrophysics with the Square Kilometre Array\\
June 8-13, 2014\\
Giardini Naxos, Italy}

\begin{document}

\section{Introduction: The Science of Globular Cluster Pulsars}

Compared with the Galactic field, globular clusters (GCs) contain large numbers of millisecond pulsars (MSPs) per unit stellar mass --- for recent reviews see: \citet{Camilo2005,Ransom2008,Freire2013}. The stellar densities in the cores of GCs are orders-of-magnitude 
larger than in the Galactic plane; these conditions create a unique environment for stellar collisions and interactions \citep[e.g.][]{Sigurdsson1993, Sigurdsson1995}.  As a result, neutron stars can swap in new companion stars and form exotic binary (and trinary) systems \citep[e.g.][]{Thorsett1999, Sigurdsson2003}, in which the neutron star is often spun-up to become an MSP through a period of mass and angular momentum transfer \citep{Alpar1982,Radhakrishnan1982}.  Furthermore, GCs have half-light radii of only a few arc-minutes or less, making them easy targets for directed searches.  Deep searches of the Galactic GC system have thus far discovered 144 pulsars in 28 clusters\footnote{See http://www.naic.edu/$\sim$pfreire/GCpsr.html for an up to date catalog.}. This includes many remarkable systems, e.g.: the fastest-spinning neutron star known \citep{Hessels2006}, potentially the most massive neutron star known \citep{Freire2008a,Freire2008b}, a double neutron star system \citep{Prince1991}, a triple system with a white dwarf and Jupiter-mass companion \citep{Thorsett1999, Sigurdsson2003}, as well as many highly eccentric binaries \citep{Ransom2005, Freire2004, Freire2008a, Lynch2012} and systems with `odd' orbits or stellar companions \citep{Lyne1993, D'Amico2001}.

Population simulations, based on the results of deep pulsar searches in the past decade, show that the Galactic GC system harbors many more pulsars still to be discovered \citep{Bagchi2011, Chennamangalam2013, Turk2013}. In particular, once extrapolated to the total sample of 157 known GCs in the Milky Way, the most recent study \citep{Turk2013} predicts a population range of potentially observable pulsars (i.e. those beamed towards us) between $600 - 3700$ (95\% confidence level). The distribution of this population among the various known sub-classes of pulsars (isolated, binaries, trinaries, eclipsing, etc.) has not been carefully explored \citep[though see][]{Verbunt2014}.  Nonetheless, past experience indicates that any significant growth in the known population inevitably leads to the discovery of some exotic and rare new kinds of pulsar systems \citep[e.g. the pulsar triple system PSR J0337+1715;][]{Ransom2014}.  In this respect, pulsars in GCs are of special interest because the peculiar environment of the core of a GC provides the conditions for producing systems that are unlike anything that can be easily formed in the Galactic field \citep[e.g.][]{Freire2004}, thus opening the possibility of detecting `Holy Grail' systems, like the first-ever MSP + black hole binary or even an MSP + MSP binary. The known GC pulsars, and those that the SKA will discover, can be studied very efficiently, because they are all located close to each other on the sky.  For example, a single observation using no more than a half dozen tied-array beams could time all 34 known pulsars in Terzan~5.\\

\noindent The science of GC pulsars falls into two main themes: \\

\noindent {\bf i) Individual systems that can be used for investigations in fundamental physics, like the equation-of-state of neutron-rich matter at super-nuclear densities and/or tests of gravitational theories}  \citep{Ransom2005, Freire2008a, Freire2008b, Demorest2010, Kramer2006} --- see also other relevant chapters in these proceedings \citep{Keane2014,Shao2014,Watts2014}. Such systems can also provide novel insights for understanding stellar evolution and accretion physics \citep{Archibald2009,Papitto2013,Bassa2014,Stappers2014}. Although these science cases can also be addressed using pulsars in the Galactic field, the radically different environment of a GC favors the formation of exotic objects, the peculiar parameters of which may be even better suited for certain applications or constraining physical models.\\

{\it Ultra-Fast Rotators}: GCs harbor some of the fastest-spinning neutron stars known, including the 716-Hz record holder in Terzan 5 \citep{Hessels2006}.  Though such sources are rare \citep{Hessels2007}, doubling or tripling the known population provides great prospects for pushing towards even faster rotation rates --- perhaps even a {\it sub}-millisecond pulsar, which could strongly constrain the maximum neutron star radius and hence the equation-of-state \citep{Hessels2006, Lattimer2001}. Unlike in the Galactic field, an MSP in a GC can {\it in principle} experience multiple episodes of recycling (due to dynamical encounters) and hence could be more effectively spun up to its limiting rotational period (as well as growing in mass). This remains to be shown directly, however.

{\it Neutron Star Masses}: Many (at least 23) highly eccentric ($e > 0.01$) pulsar binary systems exist in GCs. For these, the neutron star mass can be constrained with the help of periastron precession \citep[e.g.][]{Freire2008a}. In the Galactic field, MSP orbits are almost exclusively extremely circular ($e < 10^{-4}$), and prohibit such measurements in most cases. Besides the obvious implications of the highest-mass sources for constraining the neutron star equation-of-state \citep{Antoniadis2013, Demorest2010}, mapping the full MSP mass distribution is an important probe of their formation in supernovae, and their later `recycling' through accretion.  Once enough NS masses have been measured, the maximum NS mass should become apparent. 

{\it Evolution and Accretion}: Some individual systems also pose interesting evolutionary puzzles, which touch on accretion physics, accretion-induced collapse, and stellar exchanges. For example, a handful of slow, apparently young pulsars are also known in GCs \citep{Lyne1996, Boyles2011}, providing interesting cases for exploring alternatives to the typical core-collapse supernova channel for forming neutron stars. The first MSP with a bloated main-sequence companion was found in a GC \citep{D'Amico2001}. This is now the prototype for a known population of a dozen such sources, some of which are seen to switch between radio MSP and low-mass X-ray binaries in quiescence and/or outburst \citep{Archibald2009, Papitto2013, Bassa2014, Stappers2014}.  In fact, there are now 21 eclipsing pulsars (so-called `black widows' and `redbacks') in GCs, meaning that there are excellent prospects for studying more such transitions in the future.

{\it Multi-wavelength Studies}: Deep multi-wavelength observations of GCs in X-rays \citep[primarily using {\it Chandra}; see e.g.][]{Grindlay2001, Pooley2002, Grindlay2002, Heinke2003a, Heinke2003b, Heinke2005, Heinke2006, Bogdanov2006, Elsner2008, Bogdanov2011}, $\gamma$-rays \citep[primarily using {\it Fermi}; see e.g.][]{Freire2011b}, and optical \citep[primarily using {\it HST}; see e.g.][]{Edmonds2001, Bassa2003, Bassa2004, Pallanca2010, Pallanca2013, Pallanca2014} provide complementary information on either the pulsar's magnetospheric emission, intra-binary emission (perhaps from a shock), or the companion itself. The next generation of extremely large telescopes (ELTs) may provide large numbers of radial velocities for constraining the mass ratios of the stellar components, as in \citet{Cocozza2006}.  Conversely, identifying high-energy GC sources as radio pulsars helps better understand the zoo of objects that can be created (e.g. cataclysmic variables, low-mass X-ray binaries, etc.) and hence the stellar evolution history of the cluster itself \citep{Heggie2008}. For GC pulsars we have an independent measure of the distance (and often the reddening), which can allow stronger constraints on several pulsar+companion quantities measured in optical, X-rays or $\gamma$-rays. The {\it Fermi} satellite is revealing $\gamma$-ray emission from several GCs \cite[e.g.][]{Abdo2010}. Comparing the total $\gamma$-ray emission with the number and properties of the known radio MSPs in the largest possible sample of GCs will be a new tool for investigating the as yet poorly understood high-energy emission mechanisms of the MSPs \citep[e.g.][]{Harding2005}, as well as constraining the MSP radio beaming factor (i.e. the fraction of the sky swept by the MSP radio beam). {\it Fermi} is not only detecting several GCs, but has started detecting individual pulsars, like PSR B1820$-$30A \citep{Freire2011b} and PSR B1821$-$24A \citep{Johnson2013}.  \\

\noindent {\bf ii) Using the cumulative population of pulsars to probe the structure, proper motion, dynamical status, magnetic field, and gas content of the cluster itself} \citep{Phinney1993, Meylan1997, Pooley2003}. Pulsar-determined quantities can constrain the still largely obscure GC evolution, as well as its relation with the history of the Galaxy.  For example, the pulsar-abundant cluster Terzan~5 contains at least two stellar populations of differing ages, and is possibly the pristine remnant of a building block of the Galactic bulge \citep{Ferraro2009}.  The same advances in computing that will allow the creation of the SKA will also allow advances in numerical simulations of GCs. Simulations of the stellar and dynamical evolution of stars in a GC will predict the observable population of neutron stars in each cluster, and this can be directly tested against observations. Neutron stars, as some of the heaviest objects in GCs, play a vital role by acting as an energy reservoir to counter the gravitational collapse through the formation and disruption of binaries. The SKA will provide a census by looking for those neutron stars seen as pulsars, and their spin and binary properties will give important clues to the dynamical state of the cluster \citep[e.g.][]{Verbunt2014}.  \\

{\it Cluster Potential}: Acceleration in the cluster potential affects the pulsars' spin derivatives and binary orbital period derivatives, which can be used to probe the cluster potential in a very direct and unique way, possibly revealing the presence of an intermediate mass black hole at the core. Strong constraints can also be set on the inner mass-to-light ratio by finding a very centrally located pulsar \citep{D'Amico2002}. Thus far, there is no certified intermediate-mass black hole (IMBH) known to exist, anywhere in the Universe. Finding such an object would lead to a wealth of studies to understand its formation and characteristics, e.g. testing if the velocity dispersion of nearby stars in the GC is related to the black hole mass, as is seen in galaxies over a few orders-of-magnitude in mass. The discovery of an MSP orbiting an IMBH would give a unique chance to directly measure the black hole spin \citep[][see also Liu 2012, PhD, U. of Manchester]{Liu2014}.

{\it Proper Motion}: The ensemble of pulsar proper motions, measured through pulsar timing, can be used to determine the cluster's proper motion and hence to infer its orbit in the Galactic gravitational potential (to some extent optical studies already measure GC proper motions, although not very precisely; though this will change with GAIA). While radio can provide accurate proper motion, optical instruments (especially 30-meter-class telescopes) can provide accurate radial motion. The combination produces a 3D velocity vector for the GC. When this is done for a sizable number of GCs, it will be possible to constrain the Galaxy's gravitational potential well. For nearby clusters, there is also the chance to detect the peculiar motions of the MSPs within each cluster. If they are in binary systems, then also the optical radial motion and hence the 3D motion of the population of binary MSPs in a GC can be determined. Again this will be a handle for constraining the GC potential well, especially in the core, where the relatively massive MSPs (compared to the mean stellar mass) typically reside.

{\it Intra-cluster Medium}: Differential dispersion and rotation measures (DMs and RMs) between the detected pulsars can be used as highly sensitive, unique probes of the intra-cluster ionized medium and magnetic field \citep{Freire2001b}. This can teach us about the stellar winds of old stars releasing plasma in the cluster and/or the interactions between this plasma and the winds of MSPs and hot stars. Furthermore, the existence of this plasma, together with sensitive X-ray limits on accretion luminosity from a central source, can be used to place an upper limit on the mass of a possible central black hole. As to the putative magnetic fields in GCs, these will be an additional ingredient in attacking the difficult problem of justifying the occurrence of large-scale magnetic fields in the Universe.

{\it Interaction History}: The types of pulsars that are found --- e.g. their spin-period distribution, locations in the cluster, and the fraction of binary versus isolated pulsars --- can vary quite drastically from cluster to cluster, and encodes information about the dynamical history of the GC \citep{Verbunt2014}. Here we will learn about the still largely unknown stages of a GC's evolution, with particular emphasis on the process leading to core-collapse.  There are indications that the situation is quite different with respect to what was believed until recently; for the relevant theory and observations see \citet{Fregeau2008} and \citet{Pooley2010}, respectively.\\

There are 157 known Galactic GCs \citep{Harris1996}\footnote{Last revision from December 2010 available at http://physwww.mcmaster.ca/$\sim$harris/mwgc.dat.}, most of which have now been searched for pulsars. Since the earliest handful of GC pulsar discoveries \citep{Lyne1987, Lyne1988, Manchester1990, Anderson1990, Manchester1991, Kulkarni1991}, the known population has exploded in the last decade. This has been thanks to a number of large surveys with Parkes \citep[e.g.][]{Camilo2000, Ransom2001a,  Possenti2003}, Arecibo \cite[e.g.][]{Hessels2007}, and the Green Bank Telescope \citep[GBT; e.g.][]{Ransom2004, Ransom2005}. All of these surveys greatly benefitted from new, wide-band pulsar data recorders that provided big leaps in time and frequency resolution, critical for identifying the fastest rotating pulsars. The first major crops of GC pulsars were in M15 \citep{Anderson1993} and 47 Tuc \citep{Manchester1990, Manchester1991, Camilo2000, Freire2003, Freire2001a}. Deep synthesis imaging underlined the potential for discovery \citep{Fruchter2000}. For example, deep imaging of Terzan 5  \citep{Fruchter2000}  spawned great interest in that cluster, which later resulted in a record 34 known MSPs being found \citep{Ransom2005, Hessels2006}. The discovery rate has recently fallen off, as we have reached the limits of what the current generation of large single dishes can do. As we now discuss, SKA1 can provide the next boom in GC pulsar discoveries, and this can be achieved in the very early days of the telescope, even with a subset of the total collecting area.

\section{Observing Pulsars in Globular Clusters with SKA1}

{\bf The search sample:} SKA1 will provide unprecedented sensitivity for targeted GC searches in the southern skies, which is fortunately also where the majority of the most massive, densest, and hence most pulsar-rich GCs can be found, i.e. in the Galactic bulge.  Compared with the greater observing and processing challenge of an all-sky search, which will require the formation of hundreds to thousands of tied-array beams, deep targeted searches of these limited fields-of-views will provide early commissioning science by quickly finding dozens, and eventually a few hundred pulsars in these clusters (once sufficient collecting area is in place, even if only 50\% of the total, and at least a basic tied-array mode is available; we also note that even an incoherent sum of $> 50$ dishes would already provide higher sensitivity for those clusters that are currently only visible to Parkes).   Follow-up timing observations will also be facilitated by the small required field-of-view.  We know that we are currently only sampling the tip of the pulsar luminosity distribution in these predominantly distant ($\gtrsim 4$\,kpc) stellar systems. In fact, recent investigations  \citep{Bagchi2011, Chennamangalam2013} confirmed that the luminosities (defined as $L_{\nu}=Sd^2$, where S is the observed flux density at central frequency ${\nu}$ in MHz, and $d$ is the GC distance) of the pulsars observed in a GC can be reproduced as the bright tail of  either a log-normal distribution --- with parameters compatible with the luminosity functions (LFs) inferred for the pulsars in the Galactic field --- or a power-law distribution with index $\sim -1$ \citep[in agreement with earlier results by][]{McConnell2004, Hessels2007}, with the former functional form providing a slightly better match to the available data.  For both the assumed LFs, in a large range of not too weak luminosities (typically above 0.5 mJy kpc$^2$), the flux density distribution follows d$\log$N/d$\log$S$\sim 0.5-1$ for any given cluster. That implies that a large increase in sensitivity will automatically bring many new sources within the reach of detection and will also provide higher-precision studies of the (mostly very faint, $S_{\rm 1400} \sim 20$\,$\mu$Jy) sources that are currently known.  This will also enable, e.g., more mass measurements and equation-of-state constraints using already known sources.  As we outline below, SKA1 can plausibly increase the known population of 144 pulsars by a factor of two to three and, most importantly, excellent laboratories for studying dense matter physics and gravity will be discovered. In some cases, it may be possible to detect all the active radio pulsars in a given cluster, providing a unique view of the star formation history and interactions over the cluster's lifetime.  In most cases, however, the SKA2 will be needed to reach the full observable sample.

Based on population synthesis simulations \citep{Turk2013}, a conservative total of $600 - 3700$ detectable (i.e. beamed towards us) pulsars in the Galactic GC system can be inferred. This rivals the total expected population of MSPs in the Galactic field, estimated at a few tens of thousands of sources \citep{Faucher2006, Lorimer2008}. Importantly, while Galactic field pulsars are spread across the full 41,000 sq. degrees of the celestial sphere, the total area required to search all known Galactic GCs combined constitutes only $\sim 1$\,sq.\,degree! Of the 157 known Galactic GCs, there are 154/152 GCs visible to SKA1-MID/SKA1-LOW for at least 2\,hrs, assuming elevation limits of 15/30\,deg, respectively. For 8-hr integrations, there are still 141/113 clusters available for SKA1-MID/SKA1-LOW, respectively.  Of these, 28 have at least one known pulsar, and hence the DM to the cluster is also known (a fact that makes searching significantly easier).

MSPs are in general weak radio sources (phase-averaged flux densities of $S_{1400}  \lesssim 1$\,mJy), and GCs are typically at $\gtrsim 4$\,kpc.  Raw sensitivity and long dwell time is thus crucial for finding the weakest cluster pulsars, though long dwell time also adds complications, as we discuss below.  The weakest known GC pulsars have $S_{1400}  \lesssim 10$\,$\mu$Jy and were discovered in multi-hour integrations using the GBT and Arecibo with several hundred MHz of bandwidth at 1.4 or 2\,GHz.  This sets the bar that SKA1 must surpass.
 
{\bf Searching for binaries:} Greater instantaneous sensitivity is not only important for finding weaker sources, but also because more than two-thirds of the known sources are in a binary system.  Orbital motion smears the pulsar signal over multiple Fourier bins, and can make such sources undetectable unless this is corrected for \citep{Ransom2001b}.  For this reason, it is likely that the intrinsic fraction of binary MSPs in clusters is larger than the observed fraction.  Typically, binary pulsars are discovered using an `acceleration search', which approximates the Doppler shift of the signal as a constant drift in the frequency domain \citep{Ransom2001b}.  This approach is relatively computationally efficient; full searches of all Keplerian orbital parameters (even in the case of a circular orbit, where only three orbital parameters are needed) are currently not tractable for general use.  Acceleration searches are only valid when the integration time of the full observation, $T_{\rm obs}$, is less than 10\% of the orbital period, $P_{\rm orb}$.  In other words, one cannot necessarily gain sensitivity to binary pulsar systems simply by integrating for a longer time \citep[though some analytical techniques do exist; see][]{Ransom2001b}.  The shortest known orbital period of a GC pulsar is $P_{\rm orb} = 1.5$\,hr.  Such a system can thus be found in a linear acceleration search of a $10$-min data set.  In the following, we therefore use 10\,min as a fiducial integration time for searches for binary pulsars in compact orbits, while we use 2\,hr as an appropriate integration time for longer-period binaries or isolated pulsars.
The assumption of a constant acceleration  is useful even for eccentric systems, although searches over a `jerk' parameter may be helpful \citep{Bagchi2013}.  With the constant-acceleration assumption, highly eccentric systems are in fact more easily detectable at most orbital phases than low-eccentricity systems\footnote{See Madsen 2013, MSc, UBC; https://circle.ubc.ca/handle/2429/44897}.

{\bf Propagation effects:} As in all pulsar surveys, interstellar propagation effects can strongly limit detectability, especially at the shortest spin periods, where residual smearing in time due to uncorrected dispersive or scattering delay can broaden the pulse in time to the point where it is no longer detectable. Many of the densest, most massive GCs known are located in the Galactic bulge. The line-of-sight DM is often large ($\gtrsim 200$\,pc\,cm$^{-3}$), as is the expected scattering. To mitigate these effects, one can observe at higher radio frequency. Taking the competing effect of the typically steep spectra of pulsars \citep[$S \propto \nu^{-\alpha}$, where $1 < \alpha < 3$;][]{Maron2000, Bates2013} into account, it turns out that the $1.4 - 2.0$\,GHz band is well suited for searching GCs with DMs greater than $\sim 100$\,pc\,cm$^{-3}$. Roughly 100 of the SKA-visible clusters have expected DM$ > 100$\,pc\,cm$^{-3}$ \citep[according to the NE2001 model of][]{Cordes2002}, and are excellent targets for SKA1-MID, which provides the maximum instantaneous sensitivity in the $1.4 - 2.0$\,GHz band. For the remaining $\sim 60$ SKA-visible clusters with lower expected DMs, SKA1-LOW (and/or SKA1-MID at 800\,MHz in some cases) presents an exciting opportunity, as we outline below.

{\bf Array configuration:} GC pulsar searches require the SKA's tied-array observing mode because the data must be recorded with at least $\sim 50$\,$\mu$s time resolution. Maximum possible sensitivity can be achieved if the data are coherently dedispersed at the average cluster DM (in the case of GCs with known pulsars) or at least at a few trial DMs for clusters with no known sources. In practice, this will likely require the VLBI-like ability to record raw voltages from the tied-array.  An important consideration is the sky area that must be covered for such searches. The known Galactic GCs typically have core radii of a few arc-seconds to one or two arc-minutes and half-light radii of $\sim 0.5 - 3$ arc-minutes  \citep{Harris1996}. GC pulsars congregate to a large extent towards the cluster core, and almost all of the known GC pulsars are within one arc-minute of the cluster's optical center-of-light. The 400-m (radius) core of SKA1-MID will provide a tied-array beam with FWHM of 0.9 arc-minute at 2 GHz, meaning that only a single beam will be needed to detect and discover the majority of the GC pulsars using this setup. In order to catch the minority of pulsars further from the core, which are still very interesting for understanding the dynamical history of the cluster, a more than sufficient coverage can be achieved with hexagonally packed grids of 7 or 13 beams. For the 600-m (radius) core of SKA1-LOW, a single 3 arc-minute tied-array beam at 350\,MHz will be sufficient. Thus, searching for GC pulsars is an excellent early science and commissioning project, which will naturally lead the way to an all-sky pulsar survey requiring hundreds to thousands of tied-array beams. More ambitious searches using a larger fraction of the SKA1-MID/LOW antennas will be more challenging. For example, to double the instantaneous sensitivity provided by SKA1-MID requires going to a radius of 4000 m from the array center, and thus 100 times more beams to cover the same area of sky. This approach is likely more interesting for follow-up than for initial searches. For example, once new pulsars are found and sufficiently well localized, the further timing observations that are needed to extract the science (e.g. precision astrometry, orbital elements, proper motion, post-Keplerian effects, etc.) can be done by using one narrow, {\it full-array} tied-array beam on each source.

{\bf Achievable sensitivity:} The 400-m (radius) core of SKA1-MID, operating from $1.4-2.0$\,GHz can achieve an rms noise of 4.0/1.2 $\mu$Jy for 10-min/2-hr integrations. Using a digital beam-former and back-end similar to those sketched in the SKA Baseline Design (i.e. capable of keeping the smearing effects of interstellar dispersion smaller than the adopted sampling time of 50 $\mu$s for
all DMs $< 1000$\,pc\,cm$^{-3}$), that rms noise translates into a limiting sensitivity (at ${\rm S/N} = 8$) of  $\sim 12/4$ $\mu$Jy for a recycled pulsar spinning at a $\sim 1$ ms period and having an intrinsic pulse duty cycle of $10\%$.  Those sensitivity limits halve for the GCs where 8hr-long tracks with SKA1-MID are possible (this fortunately includes some of the most promising targets)\footnote{It is important to note that longer integrations do not necessarily improve sensitivity to binary pulsars because only a coherent search of the possible orbital parameters can fully recover the Doppler-smeared signal.  Nonetheless, by the time that the SKA is operational, it may be possible to use more advanced binary search techniques than those typically used at present.}.  
SKA1-MID can reach a factor of $\sim 3$ deeper than the state-of-the-art searches performed using the GBT and the GUPPI  back-end \citep{DuPlain2008}.  Only 32 clusters within declination, $\delta$, of $-2$\,deg$ < \delta < +38$\,deg are visible to Arecibo.  For most of the $30$ clusters south of $\delta = -40$\,deg the best limits have been reached by Parkes; SKA1-MID can go $\sim 11$ times deeper in these cases.  The 600-m (radius) core of SKA1-LOW, operating from $250-450$\,MHz, can achieve $\sim 100/30$\,$\mu$Jy sensitivity for 10-min/2-hr integrations (depending on the line-of-sight; for clusters in the Galactic bulge, $T_{sky}$ will be significant).  For comparison, the all-sky GBNCC survey at 350\,MHz achieves $\sim 1$\,mJy sensitivity \citep{Stovall2014}.  

\begin{figure}[htbp]
\begin{center}
\centerline{\includegraphics[width=12cm]{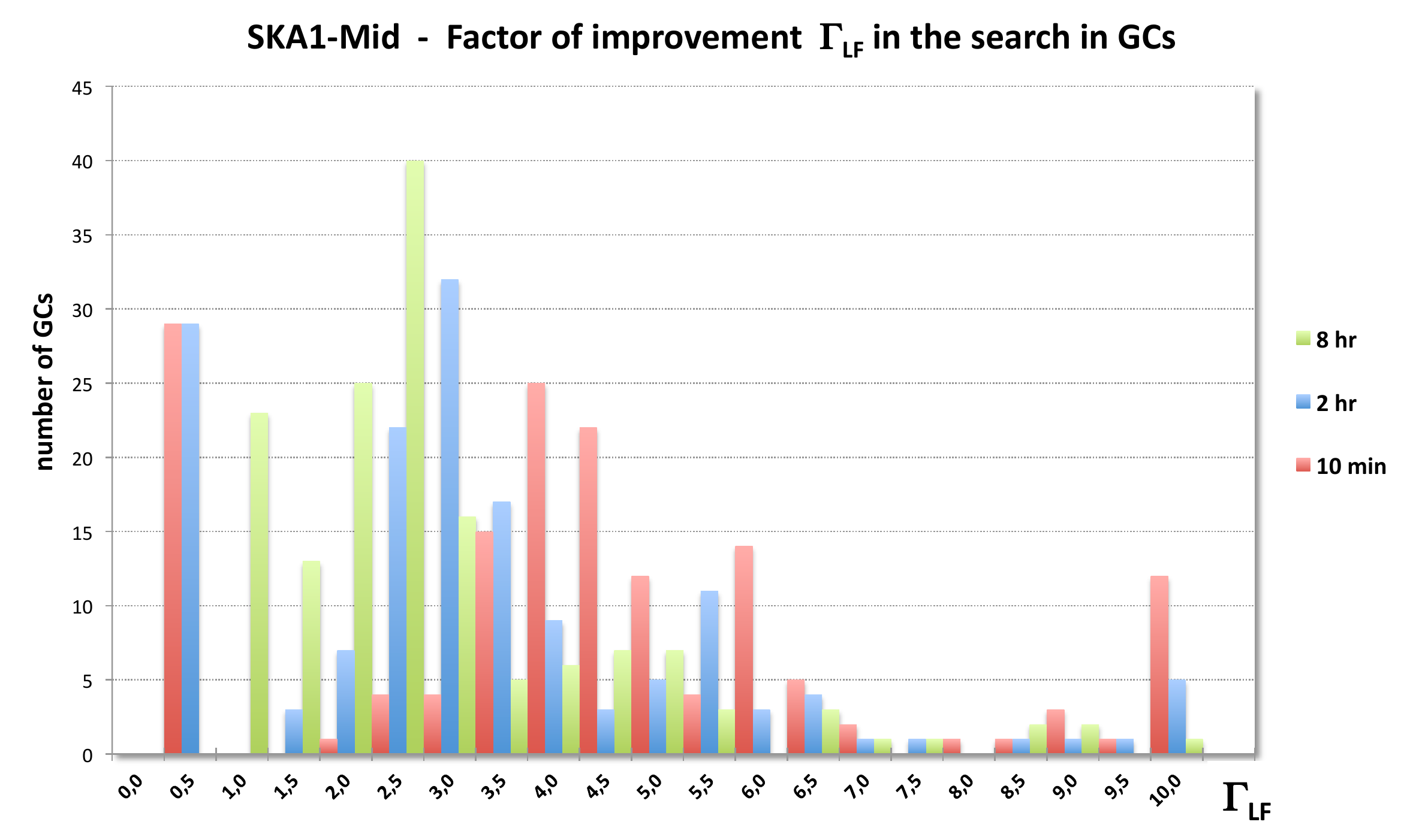}}
\caption{A histogram of the distribution of $\Gamma_{LF}$, which represents the growth of the probed area of the pulsar luminosity function, for all GCs visible to the SKA.  Here we assume use of SKA1-MID in 10-min, 2-hr, and 8-hr observations, compared to state-of-the-art observations with Arecibo, GBT, and Parkes.  See text for further details.}
\label{default}
\end{center}
\end{figure}

{\bf Available discovery space:} Figure~1 illustrates the discovery space that is open to SKA1-MID for these kind of pulsar searches (a similar analysis can be made for SKA1-LOW).  We define the quantity $\Gamma_{LF}$ to represent the growth of the probed area of the pulsar luminosity function (LF) of a given GC.  This assumes a survey performed with SKA1-MID at 1700\,MHz central frequency, with 600\,MHz bandwidth and other survey parameters like those in the SKA Baseline Design. In particular, the histogram of Fig.1 is obtained by comparing the sensitivity of this SKA1-MID survey with that of the best possible GC searches already carried out or still ongoing.  For each cluster, we assume that the best-possible search can be conducted by either Arecibo, Parkes, or GBT, depending on the declination of the cluster (Arecibo is the best telescope for all clusters within $-2$\,deg$ < \delta < +38$\,deg, Parkes is the best telescope for all clusters $\delta \lesssim -40$\,deg, and the GBT is the best available for $\delta \gtrsim -40$\,deg and outside the Arecibo declination range).  For the purpose of comparison, all flux density limits have been scaled to 1400\,MHz using a pulsar spectral index $\alpha = -1.7$.  We also assume, for simplicity, that the back-ends used in these various surveys are equally as good at minimizing the effects of dispersion.  Finally, also the effects of scattering in the ISM have been assumed to be the same at the central frequencies of SKA1-MID and at the reference frequencies of the other telescopes ($\sim 1400$\,MHz for Parkes and Arecibo, and  $\sim 2000$\,MHz for GBT). Using these assumptions, and an effective pulsar duty cycle of 25\% (which is a typical for millisecond pulsars), we calculated survey flux density limits. The next step in producing Fig.~1 was to feed the calculated sensitivity limits to a log-normal pulsar LF with mean (in units of mJy\,kpc$^2$ expressed in a logarithmic base-10 scale) of $-1.1$ and dispersion 0.9, which is known to reproduce the observed data \citep{Bagchi2011,Chennamangalam2013}.

We considered the distributions of $\Gamma_{LF}$ for 3 hypothetical surveys looking at each GC for 10 min, 2 hr, and 8 hr (color-coded in red, blue and green), respectively. It is evident that, for most GCs, SKA1-MID provides up to a factor 5 improvement in probing the LF. The average $\Gamma_{LF}$ over the entire population is between 3 and 4\footnote{Power-law LFs (with index between $-1.0$ and $-0.7$, and cut-off luminosity in the confidence range $0.1-0.5$\,mJy\,kpc$^2$) were also explored. Although the shape of the histogram distribution of $\Gamma_{LF}$ is different with respect to the adoption of a log-normal function, the average value of $\Gamma_{LF}$ over the whole GC population is similar.}.   The higher average $\Gamma_{LF}$ for shorter integration times is due to the shape of the log-normal LF, which is steeper at higher luminosity.  In other words, a given increment in sensitivity produces higher values of $\Gamma_{LF}$ when only the brightest part of the LF is sampled, which typically occurs for short pointings.  This effect also largely explains the second peak at $\Gamma_{LF}\sim 10$, which results from the $\sim 10$ very distant GCs that are visible from Parkes. The increment in sensitivity of SKA1-MID with respect to Parkes is large and, moreover, if a pulsar should be discovered in those GCs by a shallow survey, its luminosity will have to be very high and thus $\Gamma_{LF}$ is correspondingly very high. Finally, Fig.1 shows that also for the $\sim 30$ targets in the visibility range of Arecibo (appearing at $\Gamma_{LF}$ between 0.5 and 1.0 for the case of short or intermediate-duration pointings), SKA1-MID can go deeper when using 8-hr integrations (Arecibo can typically track sources for only $1-3$\,hrs).

We note that a direct extrapolation from $\Gamma_{LF}$ to the expected number of pulsar discoveries in each GC is impossible for the majority of clusters, where there is currently no known pulsar.  Such predictions are also uncertain for the GCs with known pulsars, because various effects (e.g. scintillation,  fraction of binaries, Doppler smearing of the pulsar signal) can bias any simple relation between $\Gamma_{LF}$ and the number of known pulsars.  Nonetheless, despite all these caveats, the number of pulsars resulting from the direct multiplication of  $\Gamma_{LF}$ with the number of already known objects is compatible with the results of sophisticated statistical analyses \citep[e.g.][]{Bagchi2011, Chennamangalam2013} applied to some of the most populated GCs.  These predict  that the number of potentially observable pulsars should be (within 95\% credible intervals) $82 - 259$ in Terzan 5, $48 - 137$ in 47 Tuc, and $48 - 191$ in M28.  In summary, $\Gamma_{LF}$ can be only taken as a figure-of-merit for the capabilities of SKA1-MID in the search for GC pulsars and is mostly useful for comparison with the current state-of-the-art Arecibo, GBT, and Parkes searches.  In this respect, we note that a downgrade of SKA1-MID by a factor $50\%$ in sensitivity would significantly reduce the effectiveness of the search for pulsars in GCs.  In such a situation, it would be difficult to go significantly deeper than previous GBT searches, and impossible to go deeper than Arecibo.  Thus, with a 50\% reduction in SKA1-MID sensitivity, a significant sensitivity improvement would be limited to the GCs visible only from Parkes, i.e 1/5 of the total sample.

\begin{figure}[htbp]
\begin{center}
\centerline{\includegraphics[width=12cm]{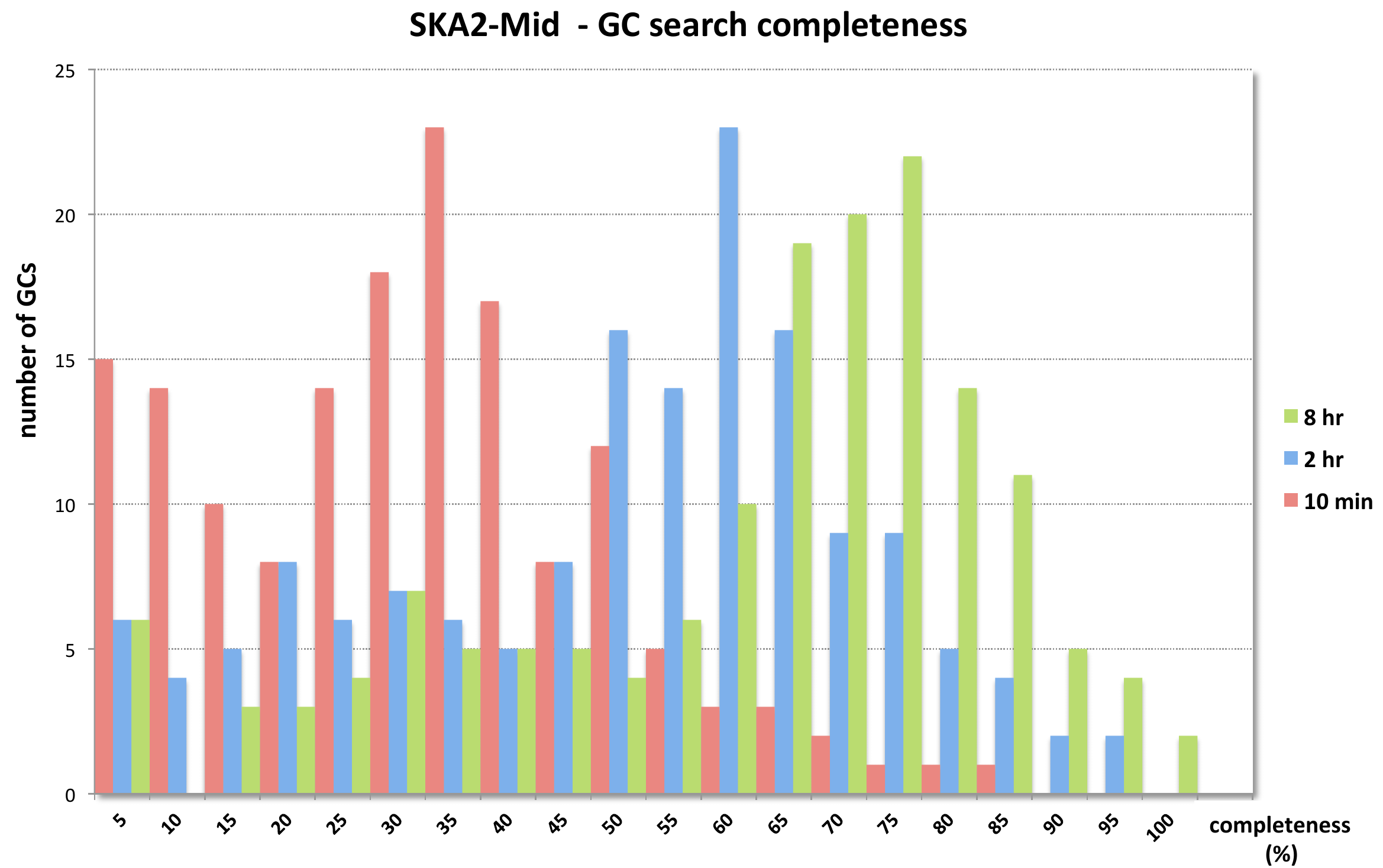}}
\caption{A histogram of the percentage completeness factors for GC pulsar searches with SKA2 (at $1700$\,MHz), using 10-min, 2-hr, and 8-hr observations.  See text for further details.}
\label{default}
\end{center}
\end{figure}

\section{Observing Pulsars in Globular Clusters with SKA2}

The great promise of SKA2 is illustrated in Fig.~2, where we show the distribution of the `completeness percentages' of a GC survey.  For example, 50\% completeness indicates the capability of probing half of the total area defined by the LF of a given GC --- i.e. half of the area under the curve is probed.  Even with 8-hr integrations, SKA1-MID cannot see all the potential pulsars in a given GC, whereas SKA2 may detect the majority or even all the active radio pulsars in a large sample of some tens of  GCs --- enabling detailed comparisons of their evolutions and structures, as well as an unbiased determination of the shape and cut-off of the LF of the millisecond pulsar population. We also note that, when SKA2 will be operational, SKA1-MID will have likely detected at least one pulsar in most GCs, which will provide the possibility for SKA2 to speed up the search for additional pulsars (and significantly reduce its computational cost) by using an already known approximate value of the DM.

\section{Conclusions}

Once constructed, SKA1-MID and SKA1-LOW will be the premier search machines for pulsars in GCs.  Their raw sensitivity will surpass that of the GBT by a factor of about 3 and 4, respectively, which opens a large discovery space for faint sources outside of the Arecibo declination range.  Though the uncertainties are large, we estimate that a doubling or tripling of the known pulsar population is possible.  In other words, anywhere from $100-300$ pulsars can be found in a full census of $\sim 150$ GCs using 10-min integrations and only one tied-array beam synthesized from the SKA Core.  This offers a great opportunity for early SKA pulsar science, even before all the collecting area is in place.  A doubling or tripling of the known GC pulsar population means finding more exceptional pulsar systems for testing strong gravity, dense matter, and more.  For example, only 3 of the known GC pulsars (all in Terzan 5) have spin rates greater than 500\,Hz.  A doubling or tripling of the known population gives good prospects for breaking the current 716-Hz rotation record held by Ter5ad.  Besides the study of individual pulsar `jewels'  the ensemble of detected pulsars provide a unique probe of the dynamics and evolution.  While SKA1 can significantly push such studies forward, SKA2 can bring us within a factor of a few of finding all the detectable (beamed towards us) pulsars in Galactic GC system, a population that has been estimated at $600 - 3700$, and which is comparable to the total detectable Galactic field population.

\section*{Acknowledgements}

J.W.T.H. acknowledges funding from an NWO Vidi fellowship and ERC Starting Grant ``DRAGNET" (337062).

\bibliographystyle{apj_short_etal}
\bibliography{Chapt_GC_PSR_SKA}

\end{document}